Preprint

# Digital requirements engineering with an INCOSE-derived SysML meta-model

James S. Wheaton[*], Daniel R. Herber

*Department of Systems Engineering, Colorado State University, Fort Collins, CO 80523*

**Abstract**

Traditional requirements engineering tools do not readily access the SysML-defined system architecture model, often resulting in ad-hoc duplication of model elements that lacks the connectivity and expressive detail possible in a SysML-defined model. Without that model connectivity, requirement quality can suffer due to imprecision and inconsistent terminology, frustrating communication during system development. Further integration of requirements engineering activities with MBSE contributes to the Authoritative Source of Truth while facilitating deep access to system architecture model elements for V&V activities. The Model-Based Structured Requirement SysML Profile was extended to comply with the INCOSE *Guide to Writing Requirements* updated in 2023 while conforming to the ISO/IEC/IEEE 29148 standard requirement statement templates. Rules, Characteristics, and Attributes were defined in SysML according to the *Guide* to facilitate requirements definition and requirements V&V. The resulting SysML Profile was applied in two system architecture models at NASA Jet Propulsion Laboratory, allowing us to explore its applicability and value in real-world project environments. Initial results indicate that INCOSE-derived Model-Based Structured Requirements may rapidly improve requirement expression quality while complementing the *NASA Systems Engineering Handbook* checklist and guidance, but typical requirement management activities still have challenges related to automation and support with the system architecture modeling software.

## 1. Introduction

As space systems grow in complexity, the demand for more cost-efficient system development programs grows in turn. As an example, the NASA-ESA Mars Sample Return (MSR) mission is "an ambitious and complex space system engineering endeavor" (Sundararajan, 2022) with multiple interfacing space systems necessary to coordinate the safe return of Martian gas and solid core samples for further study on Earth. The MSR mission has recently undergone its second Independent Review Board assessment that includes a probable program life cycle cost estimate of $8-11 billion, "strong irrefutable evidence" that strong systems engineering is a crucial factor for mission success, and recommendations to refactor the program architecture to control costs (Figueroa et al., 2023). Strong systems engineering is increasingly model-based, and digital requirements engineering aims to improve quality and reduce cost in support of the organization's digital engineering goals.

[*] Corresponding author. *E-mail address:* james.wheaton@colostate.edu

### *1.1. Digital Engineering (DE)*

Recognition of the current and potential impact of digital models, including those used in Model-Based Systems Engineering (MBSE), has led to the development by the US Department of Defense of a strategy for taking greater advantage of digital models to transform the system development process. DE is "an integrated digital approach that uses authoritative sources of system data and models as a continuum across disciplines to support lifecycle activities from concept through disposal" (Office of the Deputy Assistant Secretary of Defense for Systems Engineering, 2018). DE is not a new discipline of engineering but rather an intentional transformation of how an organization integrates and performs its engineering activities to achieve higher quality and efficiency (Noguchi, Wheaton and Martin, 2020). In recognition of the advantages of digital workflows, NASA has published a Digital Transformation strategy (Marlowe, Haymes and Murphy, 2022), MBSE strategy (Weiland, 2021), and DE Acquisition Framework Handbook (Office of the NASA Chief Engineer, 2020). One of DE's goals is to provide an enduring Authoritative Source of Truth (ASoT) of the system to improve communication and decision-making. The system architecture model is one component of the ASoT, typically integrated in a centralized repository, and the system requirements may be created in the architecture model or synchronized with the model from a Requirement Management Tool (RMT). Digital requirements engineering further integrates requirements with the ASoT, enabling formal verification and validation (V&V) activities that may be automated to improve model confidence and ease stakeholder reviews (Duprez et al., 2023).

### *1.2. Requirements Engineering*

Requirements engineering is a subset of systems engineering that encompasses requirements development and requirements management. The ISO/IEC/IEEE 29148:2018 standard defines requirements engineering as "an interdisciplinary function that mediates between the domains of the acquirer and supplier or developer to establish and maintain the requirements to be met by the system, software or service of interest. Requirements engineering is concerned with discovering, eliciting, developing, analyzing, verifying (including verification methods and strategy), validating, communicating, documenting and managing requirements" (ISO/IEC/IEEE, 2018). The necessary and tailored range of requirements engineering activities necessitates the use of attendant attributes to organize the model-connected information of the requirement, emphasizing that the familiar "shall" statement is only one attribute of a well-managed and model-connected requirement.

The INCOSE *Guide to Writing Requirements* (GtWR) provides a current perspective of well-formed requirements, and it defines a requirement statement as "the result of a formal transformation of one or more sources, needs, or higher-level requirements into an agreed-to obligation for an entity to perform some function or possess some quality within specified constraints with acceptable risk" (Wheatcraft and Ryan, 2023). The GtWR emphasizes that the requirement statement forms the basis of contractual language, and then presents a rules-based structured format for facilitating that communication. The GtWR notes a data-centric practice using a RMT or Systems Modeling Language (SysML) tool as opposed to spreadsheets or documents to model and present requirement expressions using diagrams for stakeholder-tailored views.

Systems engineering handbooks provide another important reference for requirements engineering activities, complementing the detailed guides, manuals, and standards cited above. The INCOSE Systems Engineering Handbook - Fifth Edition provides updated Sections 2.3.5.2 and 2.3.5.3 that incorporate the latest INCOSE guides and manuals on needs and requirements engineering (INCOSE, 2023). The NASA Systems Engineering Handbook (Hirshorn, Voss and Bromley, 2017) describes the traditional NASA requirements definition and management processes in Sections 4.2 and 6.2, respectively, emphasizing bidirectional traceability, and including a checklist in Appendix C and an informal set of characteristics similar to those defined in the latest INCOSE GtWR. The INCOSE GtWR and NASA sets of characteristics are compared in Sec. 3 and found to be complementary.

### *1.3. Systems Modeling Language*

SysML is the standard language for modeling systems structure, behavior, requirements, and rules (parametrics). The Object Management Group has published SysML version 1.7 (OMG, 2022), which is expected to be the final version in the 1.x series as the standards development effort shifts focus to the new version 2. SysML v2 will not be discussed in this paper, but is expected future work as one major difference is the consistent distinction between a requirement definition and requirement usage that aids in reuse. Property-based requirements (Bernard, 2012) have previously been described and emphasized using SysML or UML modeling to reduce the ambiguity of system requirement statements, which are maintained separately and manually synchronized.

SysML provides rudimentary facilities for modeling system requirements, including the primary attributes: ID, name, and text; and the relationships: derive, refine, satisfy, verify, and trace, which is discouraged in favor of the more precise relationships. Requirement rationale and type are not attributes provided by the SysML standard but are customizations of the profile often provided by systems architecture modeling tools. According to the standard, SysML Requirements may be shown in a Requirements Diagram, or placed on other SysML diagrams to highlight relationships for certain stakeholder views; requirements tables and matrices are non-normative.

Due to the perceived inadequate facilities for modeling and managing requirements, SysML has not been favored by requirements engineers (Wheatcraft and Ryan, 2023). Its apparent advantages with respect to its graphical syntax (diagrams) have not been enough to satisfy the critical need of managing possibly thousands of requirements. However, SysML makes it possible to extend the language through Stereotypes thus availing the systems engineer with meta-modeling capabilities to define custom elements and to use them as any other SysML element. This meta-modeling capability is what enables the MBSR approach described in this paper.

### *1.4. Overview*

This paper presents an extension to the MBSR approach developed by Herber and Eftekhari-Shahroudi (2023) and Herber, Narsinghani and Eftekhari-Shahroudi (2022) that incorporates the INCOSE GtWR (Wheatcraft and Ryan, 2023) and employs the standard (ISO/IEC/IEEE, 2018) template for requirement statements using a notional NASA system to demonstrate the use of the SysML meta-model. The rest of the paper is organized as follows: Sec. 2 describes how the MBSR meta-model is defined and used; Sec. 3 discusses the advantages and limitations of this approach; and Sec. 4 provides some conclusions and summary of future work.

## 2. Model-based structured requirements definitions and usage

### *2.1. Model-based "shall" statement pattern slots*

Following Carson (2015), requirement statements with uniform structure are shown to improve quality and have been adopted and recommended by IEEE (ISO/IEC/IEEE, 2018) and INCOSE (2023). Textual "shall" statements with parts in a standard order are easier to write, parse, and verify due to the regular structure that guards against ambiguity and complex grammar. The INCOSE GtWR contains a comprehensive discussion of requirement statement patterns in Appendix C (Wheatcraft and Ryan, 2023). The ISO/IEC/IEEE 29148:2018 standard (ISO/IEC/IEEE, 2018) presents two templates with an implied "shall" after the Subject:

**[Subject] [Action] [Constraint of Action]**
*OR*
**[Condition] [Subject] [Action] [Object] [Constraint of Action]**

The Subject refers to the part of the system corresponding to the same level as the requirement. The Action signifies that the Subject *does* something and that the statement is written in the active voice, avoiding superfluous and possibly confusing verbiage such as “be capable of”. Every requirement must have a verifiable Constraint. Although every requirement has an associated condition of when it is active, the first template may be used in the high-level functional requirements when the Condition is ubiquitous (Wheatcraft and Ryan, 2023). The latter pattern is used as a default value for the redefined ‘Text’ SysML Property.

The MBSR approach developed by Herber and Eftekhari-Shahroudi (2023) and Herber, Narsinghani and Eftekhari-Shahroudi (2022) builds upon the template or pattern concept by making the pattern slots into SysML Properties of a Structured Requirement-stereotyped element, thus providing corresponding model elements and even diagrams as a model-based supplement to the textual “shall” statement (Figs. 1 and 2). Earlier meta-models of MBSR limited the property types to appropriate SysML types such as Block and Constraint Block, but this approach was later found to be too restrictive, and the standard SysML meta-model decision to use NamedElement was adopted (OMG, 2022). The MBSR meta-model works by using Generalization relationships with the existing SysML Requirement, taking advantage of the standard syntax and semantics of SysML Requirements while separating concerns of the Structured Requirement proper and the organizational requirement with attributes peculiar to the organization.

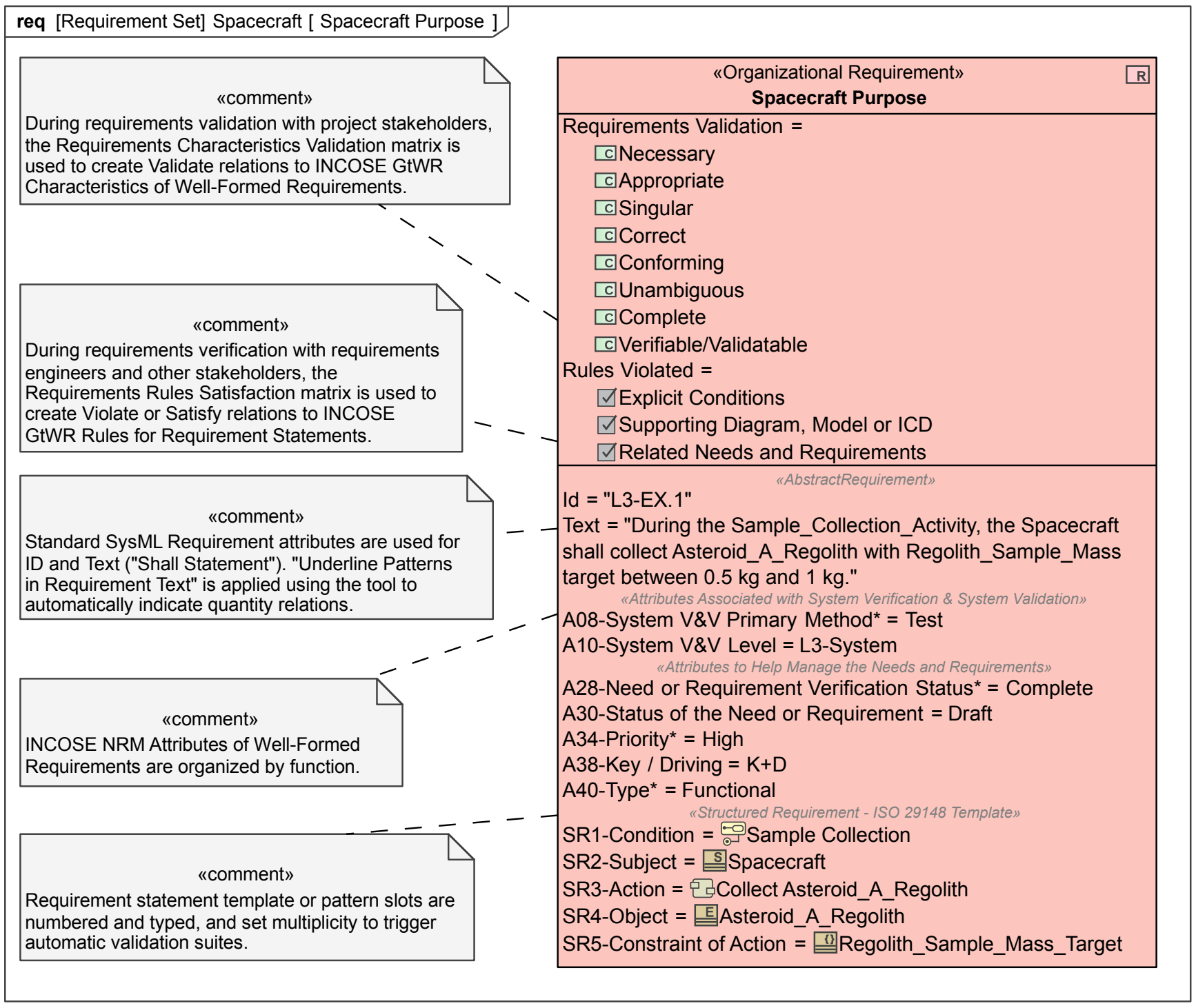

Fig. 1. Example MBSR with INCOSE GtWR Attributes, Characteristics, and Rules.

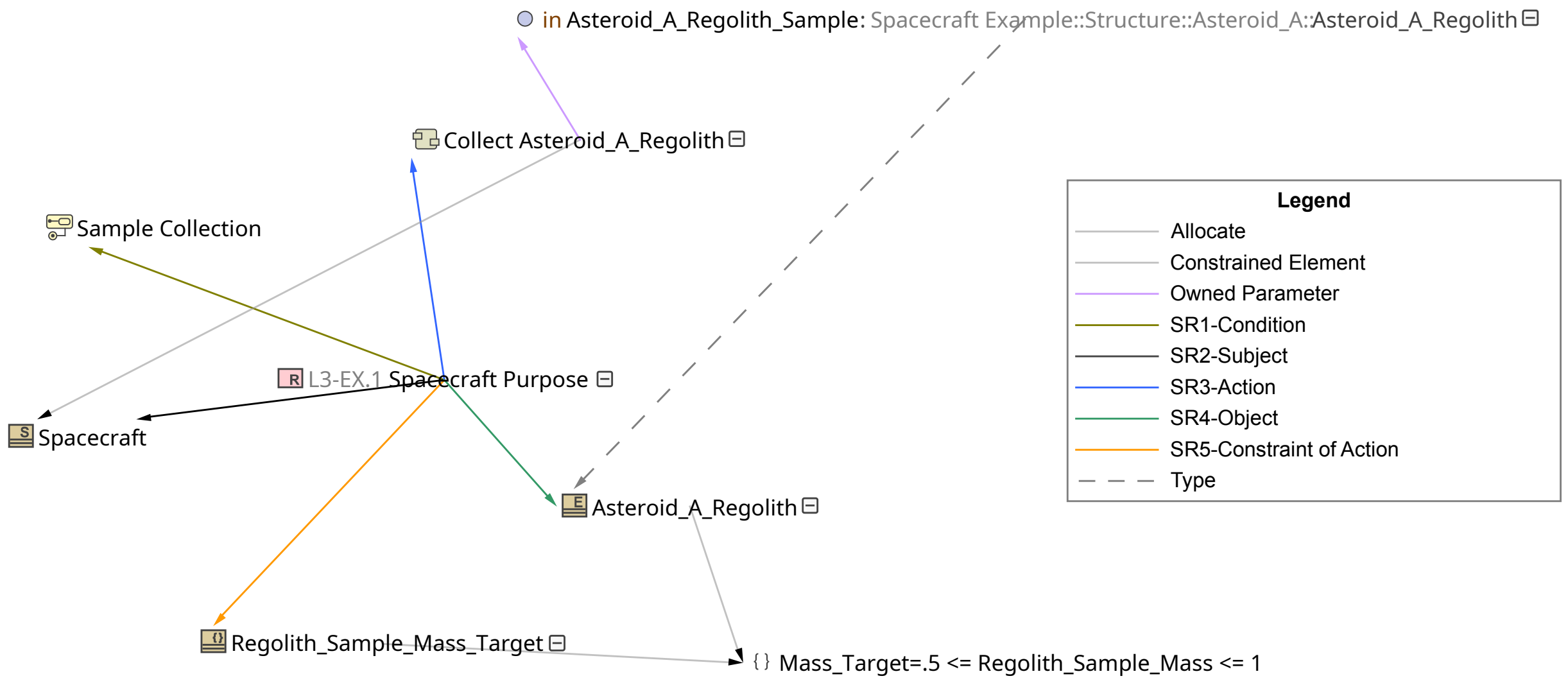

Fig. 2. Relation Map of example MBSR and related system architecture SysML elements.

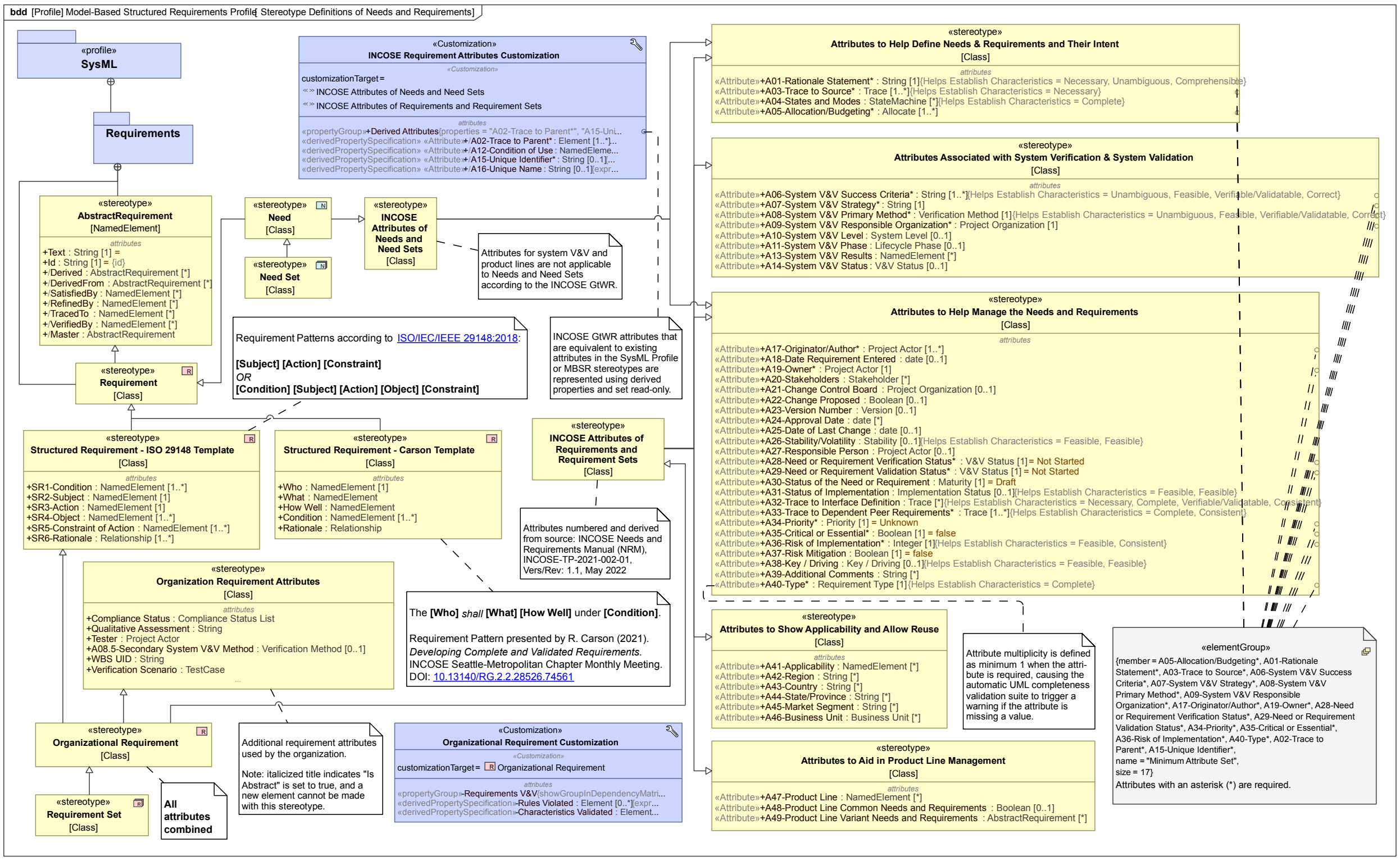

Fig. 3. SysML Stereotype definitions of INCOSE-derived MBSR using generalization relationships.

### *2.2. Attributes and rules contribute to well-formed requirement characteristics*

The MBSR extensions presented in this paper (Fig. 3) explore the utility of adding the 49 attributes and 42 rules described in the INCOSE GtWR with appropriate SysML standard and custom types defined. The Attributes are grouped according to their purpose and numbered to maintain order and to aid in searching; an elementGroup is defined to mark the attributes that are considered a minimum set by the GtWR. Notably, some Attributes are already defined elsewhere, such as A15 (ID) and A16 (Name), and are modeled using «Customization» attributes that read the standard SysML properties for completeness of the INCOSE-derived MBSR profile. A Requirement Set Stereotype is defined to distinguish from a normal Organizational Requirement, as in the GtWR, and provides additional querying, filtering, and meta-modeling capabilities for isolating Sets. Likewise, Needs and Need Sets are defined and inherit some of the same Attributes, with a different icon to visually distinguish them from Requirements.

Attributes help establish Characteristics, of which there are exactly 9 for needs and requirements, and 6 for need sets and requirement sets (Table 1). When a requirement or requirement set is verified and validated, a SysML «satisfy» relationship is added from each requirement to the respective Characteristic, providing model data and metrics of their well-formedness. The 42 Rules provided by the GtWR were also added to the MBSR meta-model. Like Attributes, Rules help establish Characteristics (Fig. A.1) of a well-formed requirement and requirement set, and during V&V activities, a «satisfy» or new «Violate» relationship is created for metrics and feedback (Fig. 4).

Table 1. Characteristics of well-formed sets and individual needs and requirements.

| ID | Name | Applicability | Derivation |
|---|---|---|---|
| C1 | Necessary | Needs and Requirements | Formal Transformation |
| C2 | Appropriate | Needs and Requirements | Formal Transformation |
| C3 | Unambiguous | Needs and Requirements | Agreed-to Obligation |
| C4 | Complete | Needs and Requirements | Agreed-to Obligation |
| C5 | Singular | Needs and Requirements | Formal Transformation |
| C6 | Feasible | Needs and Requirements | Agreed-to Obligation |
| C7 | Verifiable | Needs and Requirements | Agreed-to Obligation |
| C8 | Correct | Needs and Requirements | Formal Transformation |
| C9 | Conforming | Needs and Requirements | Formal Transformation |
| C10 | Complete | Need Sets and Requirement Sets | Formal Transformation |
| C11 | Consistent | Need Sets and Requirement Sets | Formal Transformation |
| C12 | Feasible | Need Sets and Requirement Sets | Agreed-to Obligation |
| C13 | Comprehensible | Need Sets and Requirement Sets | Agreed-to Obligation |
| C14 | Able to be validated | Need Sets and Requirement Sets | Agreed-to Obligation |
| C15 | Correct | Need Sets and Requirement Sets | Formal Transformation |

### *2.3. Using pattern structure, attributes, rules, and characteristics to write higher-quality requirements*

This MBSR SysML Profile (Herber, Narsinghani and Eftekhari-Shahroudi, 2024) was used to develop over 150 requirements and 50 requirement sets at NASA Jet Propulsion Laboratory for the Mars Returned Sample Handling project currently in Pre-Phase A. Drafts of requirements were received from the subsystem engineering leads, entered into No Magic Cameo Systems Modeler 2022x (No Magic, 2023), with given information such as rationale, verification method, verification approach, and additional comments added to respective GtWR-based Attributes. Matrix diagrams were created, as in Fig. 4, to aid in the revision of the requirement statements, providing visual feedback as a kind of checklist, and data for metrics used to triage requirements for later revision. All Rules,

Attributes, and Characteristics contain the corresponding documentation from GtWR to further assist the usage of the Profile with tooltips. To model Defined Terms as described in GtWR, Cameo Systems Modeler Glossary Tables were filled with Terms and synonyms—often acronyms—active hyperlinks to the definition source, and SysML Allocate relationships to system model elements and diagrams, providing automatic underlining of Defined Terms used consistently throughout requirements statements. In addition, other meta-model elements were added to the Profile to capture the broader range of related information: Goal, Assumption, Project Actor/Role/Organization, Requirement Type relevant to NASA/JPL, and a Second System V&V Method as Attribute A08.5.

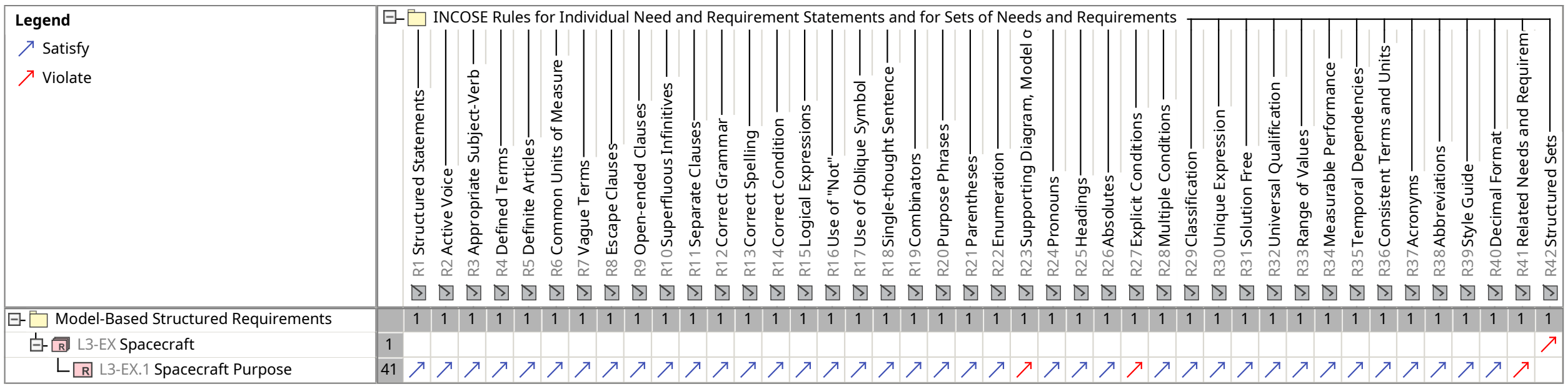

Fig. 4. A Requirements Satisfaction Matrix of INCOSE GtWR Rules, with additional Violate relationships shown in red.

## 3. Discussion of the advantages and limitations

The immediate advantage of this MBSR approach was the ability to keep architecture modeling activities confined to a single tool, and without the loss of expressiveness that would normally result from using only the standard SysML Profile. Requirements were exported to stakeholder views, including a custom PowerPoint template for assisting in reviews, an internal website (Web Report), and Excel spreadsheets (adding relevant table columns as needed). However, this approach came with its own disadvantages that became evident as limitations of the tool were encountered (Table 2).

This tool-supported MBSR Profile satisfies the NASA Systems Engineering Handbook (Hirshorn, Voss and Bromley, 2017) guidelines, specifically the Characteristics (refer to Table 1): (C3 or C9) ‘clear’, (C8) ‘correct’, (C6) ‘feasible to obtain’, (C3) ‘unambiguous in meaning’, (C14) ‘can be validated’, (C5) ‘not redundant’, (C13 and C15) ‘adequately related with respect to terms used’, and (C10 and C11) ‘not in conflict with one another’. Additionally, this MBSR Profile supports the bidirectional traceability and the creation of NASA-requested artifacts such as the Requirements Allocation Sheet, TBX report, and Requirements Verification and Validation Matrices. While the NASA Systems Engineering Handbook and INCOSE GtWR (and related guides and manuals) are complementary, the GtWR provided enumerated and defined precision amenable to SysML meta-modeling and reuse in the system architecture model.

Table 2. Advantages and disadvantages of this MBSR approach using No Magic Cameo Systems Modeler 2022x.

| Advantages | Disadvantages |
|---|---|
| Requirement IDs are customizable, with predictable increments | Requirement IDs may conflict |
| Requirement IDs may be typed in directly in diagram views | Requirement IDs are tedious to increase/decrease in bulk |
| Requirements are exportable in custom Word/Excel/PowerPoint | Requirement IDs are tedious to set with normal dialogs |
| Only used Attributes appear by default | Newly filled Attributes will appear in all affected diagrams, and require manual suppression on Requirement symbols |
| Any Attribute is displayable and sortable in a table | Duplicate Attributes seem to appear under multiple headings |
| Requirement tables may be synchronized with Excel spreadsheets | Legends cannot adorn table cells, only table rows |

| Advantages | Disadvantages |
| --- | --- |
| MBSR Attributes are quickly searchable/filterable by number, e.g. "A28" or "SR" | SysML Properties may only have one (1) Owner, preventing reuse in organization-defined Attribute Sets |
| SysML meta-modeling supports near-limitless customization | Table adorning/loading can be slow |
| Custom-query Relation Map diagrams can be made based on MBSR relationships and stakeholder intent | Writing custom Groovy scripts is challenging due to inadequate API documentation |
| Matrices with embedded tooltip documentation and double-click entries supports efficient workflows | Table scrolling performs sequential loading, temporarily revealing blank rows |
| Requirements engineering access to full system architecture model supports ASoT and precision | Reports can be slow to generate (up to an hour for 100+ diagrams) |
| Combined use of a singular tool is cost-effective for the organization | Requirements management Attributes such as Date of Last Change and Version Number require either custom scripting or manual entry |
| Consistent use of textual requirement statement patterns aids verifiability against MBSR attribute values | Defined Terms allocations to model elements may be duplicative of Structured Requirement attribute values |
| Glossary terms are underlined throughout the model, with multiple definitions visible in the tooltip | Many requirement attributes may overwhelm new users, so a minimum set should be defined by the organization |
| Collaboration plugin supports simultaneous team use | Glossary terms are sometimes randomly not underlined, never underlined in the Web Report, or the tooltip does not reliably appear |
| TBX summary table is easily made with scope, and TB[CDRN] regular expression as a filter on all fields | Mistakes made in the Shared Profile may destroy model information |
| Full ISO 80000 units of measure are available for use and extension | |

## 4. Conclusions and future work

This paper discussed an extension of prior MBSR research using the latest INCOSE Guide to Writing Requirements to improve requirements quality and connectedness contributing to an ASoT. Experience gained from applying this MBSR approach to a NASA JPL system development project was shared in brief, and the advantages and disadvantages listed may be applicable to other systems architecture modeling tools. By encoding the modelable elements from the INCOSE GtWR into the systems architecture modeling tool, rapid improvement in real-world requirements quality was achieved by a systems engineering student intern. The SysML meta-modeling demonstrated in this MBSR Profile supports a DE approach to system development and may reduce costs and improve quality if deployed at scale.

Future work may include MBSR meta-modeling using SysML v2 with particular interest in how the difficulties discussed above may be addressed with the simpler and more expressive language. The existing open-source MBSR Profile (Herber, Narsinghani and Eftekhari-Shahroudi, 2024) will continue to improve as it is tested in more types of systems and organizations, with exploration of scripted behaviors and other tool affordances/improvements for simplifying workflows and ensuring consistency of model information. SysML-based architecture modeling tools may facilitate the MBSR approach by supporting requirements management needs such as automatic timestamps, organization-defined collaborative workflows, extensible and rigorous requirement identifier definition, and enhanced speed of the software. The authors believe that MBSRs have the potential to highlight and leverage SysML strengths while satisfying requirements engineering and other stakeholder needs, but further testing, feedback, and experimentation are needed to validate this claim.

## Acknowledgments

The authors would like to thank NASA Jet Propulsion Laboratory for supporting this research. The opinions expressed herein do not necessarily represent the views of NASA or Jet Propulsion Laboratory.

# Appendix A. SysML-encoded information from the INCOSE Guide to Writing Requirements

**Legend**
↗ Allocate

Model-Based Structured Requirements Profile [Model]

| INCOSE Rules for Individual Need and Requirement State | | INCOSE Characteristics of Individual Nee | C1 Necessary | C2 Appropriate | C3 Unambiguous | C4 Complete | C5 Singular | C6 Feasible | C7 Verifiable/Validatable | C8 Correct | C9 Conforming | INCOSE Characteristics of Sets of Need a | C10 Complete | C11 Consistent | C12 Feasible | C13 Comprehensible | C14 Able to be Validated | C15 Correct |
|---|---|---|---|---|---|---|---|---|---|---|---|---|---|---|---|---|---|---|
| | | | 2 | 3 | 30 | 16 | 8 | 2 | 25 | 11 | 10 | | 4 | 10 | 3 | 8 | 7 | 7 |
| R1 Structured Statements | 5 | 5 | | | ↗ | ↗ | | | ↗ | ↗ | ↗ | | | | | | | |
| R2 Active Voice | 4 | 4 | | ↗ | ↗ | ↗ | | | ↗ | | | | | | | | | |
| R3 Appropriate Subject-Verb | 5 | 3 | | ↗ | ↗ | | | | ↗ | | | 2 | ↗ | | | | ↗ | |
| R4 Defined Terms | 6 | 2 | | | ↗ | | | | ↗ | | | 4 | | ↗ | | ↗ | ↗ | ↗ |
| R5 Definite Articles | 2 | 2 | | | ↗ | | | | ↗ | | | | | | | | | |
| R6 Common Units of Measure | 4 | 4 | | | ↗ | ↗ | | | ↗ | ↗ | | | | | | | | |
| R7 Vague Terms | 3 | 3 | | | ↗ | ↗ | | | ↗ | | | | | | | | | |
| R8 Escape Clauses | 2 | 2 | | | ↗ | | | | ↗ | | | | | | | | | |
| R9 Open-ended Clauses | 4 | 4 | | | ↗ | ↗ | ↗ | | ↗ | | | | | | | | | |
| R10 Superfluous Infinitives | 2 | 2 | | | ↗ | | | | ↗ | | | | | | | | | |
| R11 Separate Clauses | 4 | 4 | | | ↗ | ↗ | | | ↗ | ↗ | | | | | | | | |
| R12 Correct Grammar | 4 | 4 | | | ↗ | | | | ↗ | ↗ | ↗ | | | | | | | |
| R13 Correct Spelling | 2 | 2 | | | ↗ | | | | ↗ | | | | | | | | | |
| R14 Correct Condition | 2 | 2 | | | ↗ | | | | | ↗ | | | | | | | | |
| R15 Logical Expressions | 2 | 2 | | | ↗ | | | | ↗ | | | | | | | | | |
| R16 Use of "Not" | 3 | 3 | | | ↗ | | | | ↗ | ↗ | | | | | | | | |
| R17 Use of Oblique Symbol | 2 | 2 | | | ↗ | | | | ↗ | | | | | | | | | |
| R18 Single-thought Sentence | 5 | 4 | | | ↗ | | ↗ | | ↗ | | ↗ | 1 | | | | ↗ | | |
| R19 Combinators | 2 | 2 | | | ↗ | | ↗ | | | | | | | | | | | |
| R20 Purpose Phrases | 2 | 2 | ↗ | | | | ↗ | | | | | | | | | | | |
| R21 Parentheses | 1 | 1 | | | | | ↗ | | | | | | | | | | | |
| R22 Enumeration | 2 | 2 | | | ↗ | | ↗ | | | | | | | | | | | |
| R23 Supporting Diagram, Model or ICD | 3 | 3 | | | ↗ | ↗ | ↗ | | | | | | | | | | | |
| R24 Pronouns | 3 | 3 | | | ↗ | ↗ | | | ↗ | | | | | | | | | |
| R25 Headings | 1 | 1 | | | | ↗ | | | | | | | | | | | | |
| R26 Absolutes | 4 | 3 | | | | | | ↗ | ↗ | ↗ | | 1 | | | ↗ | | | |
| R27 Explicit Conditions | 3 | 3 | | | | ↗ | | | ↗ | ↗ | | | | | | | | |
| R28 Multiple Conditions | 2 | 2 | | | ↗ | | | | ↗ | | | | | | | | | |
| R29 Classification | 2 | | | | | | | | | | | 2 | ↗ | ↗ | | | | |
| R30 Unique Expression | 3 | 2 | ↗ | | | | | | | | ↗ | 1 | | ↗ | | | | |
| R31 Solution Free | 1 | 1 | | ↗ | | | | | | | | | | | | | | |
| R32 Universal Qualification | 3 | 3 | | | ↗ | | | | ↗ | ↗ | | | | | | | | |
| R33 Range of Values | 6 | 5 | | | ↗ | ↗ | | ↗ | ↗ | ↗ | | 1 | | | ↗ | | | |
| R34 Measurable Performance | 4 | 3 | | | ↗ | ↗ | | | ↗ | | | 1 | | | ↗ | | | |
| R35 Temporal Dependencies | 3 | 3 | | | ↗ | ↗ | | | ↗ | | | | | | | | | |
| R36 Consistent Terms and Units | 7 | 3 | | | ↗ | | | | | ↗ | ↗ | 4 | | ↗ | | ↗ | ↗ | ↗ |
| R37 Acronyms | 6 | 2 | | | ↗ | | | | | | ↗ | 4 | | ↗ | | ↗ | ↗ | ↗ |
| R38 Abbreviations | 5 | 1 | | | | | | | | | ↗ | 4 | | ↗ | | ↗ | ↗ | ↗ |
| R39 Style Guide | 7 | 3 | | | | ↗ | ↗ | | | | ↗ | 4 | | ↗ | | ↗ | ↗ | ↗ |
| R40 Decimal Format | 4 | 3 | | | ↗ | ↗ | | | | | ↗ | 1 | | ↗ | | | | |
| R41 Related Needs and Requirements | 6 | 2 | | | | ↗ | | | | | ↗ | 4 | ↗ | ↗ | | ↗ | | ↗ |
| R42 Structured Sets | 5 | | | | | | | | | | | 5 | ↗ | ↗ | | ↗ | ↗ | ↗ |

Fig. A.1. Allocation Matrix used as a quick reference of which INCOSE GtWR Rules contribute to which Characteristics.